\documentclass[fleqn,usenatbib]{mnras}

\usepackage{newtxtext,newtxmath}

\usepackage[T1]{fontenc}
\usepackage{ae,aecompl}

\usepackage{etoolbox}
\makeatletter
\patchcmd\@combinedblfloats{\box\@outputbox}{\unvbox\@outputbox}{}{\errmessage{\noexpand patch failed}}
\makeatother

\usepackage{graphicx}	
\usepackage{amsmath}	
\usepackage{amssymb}	
\usepackage{pdflscape}
\usepackage{xcolor}

\defcitealias{2017ApJS..230....4F}{F17}

\newcommand{\flbox}{\ensuremath{F_\mathrm{box}}}

\newcommand{\fboxsb}{\ensuremath{m_{R}^{\flbox}}}

\definecolor{2010A}{RGB}{102,194,165}
\definecolor{2010B}{RGB}{252,141,98}
\definecolor{2010C}{RGB}{141,160,203}
\definecolor{2010D}{RGB}{231,138,195}
\definecolor{2011A}{RGB}{166,216,84}
\definecolor{2012A}{RGB}{255,217,47}
\definecolor{2012B}{RGB}{229,196,148}
\definecolor{2012C}{RGB}{179,179,179}

\title[PTF SN Ia rate]{The volumetric rate of normal type Ia supernovae in the local universe discovered by the Palomar Transient Factory}

\author[C. Frohmaier et al.]{C. Frohmaier,$^{1,2}$\thanks{E-mail: chris.frohmaier@port.ac.uk}
M. Sullivan,$^{2}$
P. E. Nugent,$^{3,4}$
M. Smith,$^{2}$
G. Dimitriadis,$^{5}$
\newauthor
J. S. Bloom,$^{3}$
S. B. Cenko,$^{6,7}$
M. M. Kasliwal,$^{8}$
S. R. Kulkarni,$^{8}$
K. Maguire,$^{9}$
\newauthor
E. O. Ofek,$^{10}$
D. Poznanski,$^{11}$
and R. M. Quimby$^{12}$
\\
$^{1}$Institute of Cosmology and Gravitation, University of Portsmouth, Portsmouth, PO1 3FX, UK\\
$^{2}$Department of Physics and Astronomy, University of Southampton, Highfield, Southampton, SO17 1BJ, UK\\
$^{3}$Department of Astronomy, University of California, Berkeley, CA, 94720-3411, USA\\
$^{4}$Lawrence Berkeley National Laboratory, Berkeley, CA, 94720, USA\\
$^{5}$Department of Astronomy and Astrophysics, University of California, Santa Cruz, CA 95064, USA\\
$^{6}$Astrophysics Science Division, NASA Goddard Space Flight Center, Mail Code 661, Greenbelt, MD 20771, USA\\
$^{7}$Joint Space-Science Institute, University of Maryland, College Park, MD 20742, USA\\
$^{8}$Cahill Centre for Astrophysics, California Institute of Technology, 1200 East California Boulevard, Pasadena, CA 91125, USA\\
$^{9}$School of Mathematics and Physics, Queen's University Belfast, Belfast BT7 1NN, UK\\
$^{10}$Benoziyo Center for Astrophysics, Weizmann Institute of Science, 76100 Rehovot, Israel\\
$^{11}$School of Physics and Astronomy, Tel-Aviv University, Tel-Aviv, 69978, Israel\\
$^{12}$Department of Astronomy/Mount Laguna Observatory, San Diego State University, 5500 Campanile Drive, San Diego, CA 92812-1221, USA
}

\date{Accepted XXX. Received YYY; in original form ZZZ}

\pubyear{2018}

\begin{document}
\label{firstpage}
\pagerange{\pageref{firstpage}--\pageref{lastpage}}
\maketitle

\begin{abstract}
We present the volumetric rate of normal type Ia supernovae (SNe Ia) discovered by the Palomar Transient Factory (PTF). Using strict data-quality cuts, and considering only periods when the PTF maintained a regular cadence, PTF discovered 90 SNe Ia at $z\le0.09$ in a well-controlled sample over three years of operation (2010--2012). We use this to calculate the volumetric rate of SN Ia events by comparing this sample to simulations of hundreds of millions of SN Ia light curves produced in statistically representative realisations of the PTF survey. This quantifies the recovery efficiency of each PTF SN Ia event, and thus the relative weighting of each event. From this, the volumetric SN Ia rate was found to be $r_v=2.43\pm0.29\,\text{(stat)}_{-0.19}^{+0.33}\text{(sys)}\times10^{-5}\,\text{SNe\,yr}^{-1}\,\text{Mpc}^{-3}\, h_{70}^{3}$. This represents the most precise local measurement of the SN Ia rate. We fit a simple SN Ia delay-time distribution model, $\propto\mathrm{t}^{-\beta}$, to our PTF rate measurement combined with a literature sample of rate measurements from surveys at higher-redshifts. We find $\beta{\sim}1$, consistent with a progenitor channel governed by the gravitational in-spiral of binary white dwarfs.
\end{abstract}




\section{Introduction}

Type Ia Supernovae (SNe Ia) occur as a result of the thermonuclear explosion of a carbon-oxygen white dwarf (CO-WD) star. The exact configuration of the progenitor system is debated \citep{2014ARA&A..52..107M}, but all likely scenarios involve the interaction of the white dwarf with a second star. The two most popular scenarios are the single-degenerate (SD) channel \citep{1973ApJ...186.1007W,1982ApJ...253..798N}, where the companion star is a main sequence, sub-giant or giant star, and the double-degenerate (DD) channel \citep{1981NInfo..49....3T,1984ApJ...284..719I,1984ApJ...277..355W}, where the companion star is another white dwarf star. Almost all models involve the transfer of mass (either a steady transfer or violently) on to the primary white dwarf, resulting, by some mechanism, in a runaway thermonuclear explosion.

The various postulated progenitor scenarios can in principle operate over a differing, but wide, range of timescales following the formation of the progenitor star, from several hundred Myrs to a Hubble time. 
The duration between an instantaneous burst of star formation to the resulting SN Ia is known as the delay-time, and measuring this delay time can provide insight as to the nature of the progenitor system. While it is not possible to make such a measurement for a single object, studies of SN Ia populations and their rates can be used to construct a so-called delay-time distribution \citep[DTD; ][]{2012PASA...29..447M}, with the observed SN Ia rate then the convolution of this DTD with a particular star-formation history (SFH). Previous studies have attempted this both for galaxies \citep[e.g.][]{2006MNRAS.370..773M,2006ApJ...648..868S,2012ApJ...755...61S}, and over cosmic time \citep[e.g.][]{2004MNRAS.347..942G,2012AJ....144...59P}, where measurements of the volumetric rate of SNe Ia can constrain the DTD assuming the cosmic SFH is known. In the absence of a direct detection of a SN Ia progenitor, the DTD offers one of the most interesting constraints on the physical systems that can lead to SNe Ia.

Observations of SN Ia host galaxies have shown the intrinsic SN Ia rate can be parametrized by a combination of the total stellar mass and the star formation rate in the galaxy \citep{2006ApJ...648..868S}. A subsequent two-component approximation to the DTD can then be formed, with SNe Ia produced from both a delayed (or \lq tardy\rq) channel, and a \lq prompt\rq\ progenitor channel that traces the star formation rate (SFR). \citet{2012AJ....144...59P} combined measurements of the volumetric SN Ia rate, and considered a two-component description of the DTD. They found that while $\approx70$ per cent of SNe Ia are produced by the prompt channel when integrated over cosmic time, at $z=0$ only $\approx25$ per cent are from prompt progenitors. This naturally suggests that the dominant progenitor channel evolves with redshift. Other simple DTD parametrizations include a $t^{-1}$ power law, that has been shown to very successfully describe observed SN Ia rates \citep[e.g.][]{2012MNRAS.426.3282M, 2013MNRAS.430.1746G,2014ApJ...783...28G}, also suggesting that younger progenitors dominates the population. As we enter an era of SN Ia cosmology where systematic uncertainties dominate the error budget \citep{Betoule2014}, understanding the astrophysical origin of SNe Ia and their mix with redshift plays an increasingly important role.

Fundamental to making progress in understanding the progenitors of SNe Ia is the calculation of precise and well-understood SN Ia rates, which is the aim of this paper. The advance of robotic telescopes in the last decade has revolutionised the search for astrophysical transients. Large-area rolling surveys, such as the Palomar Transient Factory \cite[PTF;][]{2009PASP..121.1334R,2009PASP..121.1395L}, the Catalina Real-Time Transient Survey \citep[CRTS][]{2009ApJ...696..870D}, Pan-STARRS \citep{2010SPIE.7733E..0EK}, the La Silla Quest Variability Survey \citep{2013PASP..125..683B}, the All-Sky Automated Survey for Supernovae \citep[ASAS-SN;][]{2014ApJ...788...48S}, and the Asteroid Terrestrial-impact Last Alert System \citep[ATLAS;][]{2018PASP..130f4505T}, have systematically scanned the sky on timescales of minutes, days, and years in their search for transient phenomena. Surveys that photometrically identify transients are typically coupled with large spectroscopic follow-up efforts \citep[e.g.][]{2015A&A...579A..40S}, enabling the classification of objects into known classes, and the identification of interesting transients. Together, modern surveys are finding thousands of extra-galactic transients each year, and populating new regions of the time-domain phase-space \citep{2012PASA...29..482K}. Studies based on population statistics of SNe have significantly benefited from these new datasets, and in this paper we use the PTF to calculate the local universe SN Ia rate.  

The PTF was an automated optical sky survey operating at the Samuel Oschin 48 inch telescope (P48) at the Palomar Observatory from 2009--2012. The detector used gave a 7.26\,deg$^2$ field-of-view, and the survey primarily operated (83 per cent of data) in the Mould $R$-band filter, with supplementary data taken in a $g^{\prime}$ filter. Around 8000\,deg$^2$ were observed with a range of different cadences, from one to five days. PTF later transitioned to the intermediate PTF (iPTF), performing a range of different cadence experiments, and more recently the detector was upgraded to begin operations as the Zwicky Transient Facility \citep[ZTF;][]{2018ATel11266....1K,2018arXiv180210218B}. This paper concerns only data taken as part of the original PTF.

Candidates were discovered via image subtraction \citep[the pipeline is described in][]{Nugent2015}, with a machine-learning algorithm \citep{2012PASP..124.1175B, 2013MNRAS.435.1047B} assigning a \lq real-bogus\rq\ (RB) score to each candidate detected. Following this, a visual inspection of the candidates was performed by the collaboration, before entering a spectroscopic classification queue. PTF discovered more than 50,000 non-moving transients, and spectroscopically confirmed around 1,900 SNe. This led to a range of studies using large samples of common transient types \citep[e.g.,][]{2014MNRAS.444.3258M,2015ApJ...799...52W,2016ApJ...820...33R}, and new work on peculiar transients with unusual properties \citep[e.g.][]{2011ApJ...732..118S,2012ApJ...755..161K,2012Sci...337..942D,2018ApJ...858...50F,2018ApJ...860..100D,2018ApJ...855....2Q}. 

In Section~\ref{sec:infrastructure}, we introduce our previous study of the PTF survey performance that we use in this paper. We describe the fundamental principles of calculating volumetric rates from survey data. In Section~\ref{sec:sn_sample_sec} we construct our SN Ia sample, and our simulation quantifying PTF's ability to discover SNe Ia is detailed in Section~\ref{sec:survey_sim}. Our final rate result, and the method to estimate our uncertainties, is presented in Section~\ref{sec:rate_rate}. Finally, in Section~\ref{sec:analysis} we compare our measurement to others, and perform an analysis of a simple $t^{-\beta}$ power law DTD using both our rate and literature results. 

Throughout, where relevant we assume a flat $\Lambda$CDM Universe with a matter density $\Omega_M=0.3$ and a Hubble constant $H_0$=70\,km\,s$^{-1}$\,Mpc$^{-1}$, and we work in the AB photometric system \citep[][]{1983ApJ...266..713O}.

\section{Calculating rates in PTF}
\label{sec:infrastructure}

In this section, we present the general methodology on which our study is based.

The volumetric SN rate, $r_v$, is calculated as the sum of $N$ SNe that exploded within a given timespan, $\Delta T$, and fixed comoving volume, $V$. Each SN in the sum is weighted to account for the likelihood of it being detected. These weights are the inverse of the \lq detection efficiencies\rq\ ($\epsilon$) of each object, with objects that are more difficult to detect (a smaller $\epsilon$) carrying a larger weight in the sum. For each SN, $1-\epsilon$ gives the fraction of similar objects, exploding in the same search volume and duration, that would have been missed by the survey. Thus the rate can be approximated by
\begin{equation}
    r_v(z)=\frac{1}{V \Delta T}\sum_{i=1}^{N} \frac{1+z_i}{\epsilon_i},
    \label{eqn:rate}
\end{equation}
where the $(1+z)$ factor corrects for cosmological time dilation. The volume $V$ is given by
\begin{equation}
V=\frac{\Theta}{41253}\frac{4\pi}{3}\left[\frac{c}{H_0} \int_{z_1}^{z_2} \frac{dz'}{\sqrt{\Omega_M(1+z')^3+\Omega_\Lambda )}}\right ]^3\textrm{Mpc}^3,
\label{eqn:volume}
\end{equation}
\noindent
where $\Theta$ is the survey search area in deg$^2$, $z_1$ and $z_2$ are the lower and upper redshift limits over which the sample is complete and the efficiencies valid, and $c$ is the speed of light. In practice, $V$ and $\Delta T$ are reasonably easy to calculate, and thus the complexity of any SN rates analysis lies in the determination of the survey detection efficiencies.

The detection efficiency compensates for incompleteness in the SN sample. There are many reasons why a survey may miss transients: the survey design, inefficiencies in the survey detection pipelines, the observing conditions, and the transient properties. Frequently these effects are conflated, for example a low cadence survey would preferentially miss transients that evolve on rapid time scales. A careful simulation of a transient survey, accounting for real observing conditions, cadences, and using the actual survey search software, is thus a critical step in calculating the efficiencies.

In this paper, we use the simulations of PTF presented in \citet[][hereafter F17]{2017ApJS..230....4F}. This study presents the \lq single-epoch\rq\ efficiencies, giving the recovered fraction of transients as a function of various parameters. These single-epoch efficiencies were calculated by inserting fake point-sources into the real PTF data, and studying their recovery efficiency. Around 7 million point sources were added into a representative subset of PTF observations taken from 2009--2012. These images were then treated identically to the real data, and passed through the subtraction, transient identification, and machine-learning stages of the PTF pipeline. \citetalias{2017ApJS..230....4F} then studied the detection efficiency of the PTF pipeline as a function of source magnitude, host-galaxy surface brightness, the seeing (or image quality), the sky brightness, and the limiting magnitude. Multi-dimensional detection-efficiency grids were then constructed, allowing the detection efficiency of any point source in any PTF image (and on any host background) to be calculated.

We build upon this study to include models of the photometric evolution of SNe Ia, with the probability of a transient being detected on each epoch on which it is observed given by the \lq single-epoch\rq\ efficiencies. For example, transient events are observed on \textit{multiple} observing epochs, with varying brightnesses according to the transient type, and with varying observing conditions. Knowledge of that transient brightness, and the observing conditions, allows the detection probability of a transient on any \textit{individual} epoch to be determined using the single-epoch efficiencies. Combining these probabilities over all the epochs on which a SN was observed then gives the overall detection efficiency for a simulated event.

In \citetalias{2017ApJS..230....4F}, a method was described for simulating a population of SNe Ia, and we adopt this for our study. In summary, we simulate millions of SNe Ia in an artificial PTF sky, exploding randomly within our search area, at random redshifts, and on random epochs. We can then \lq observe\rq\ this population of objects and, for each artificial event, calculate whether the event would have been recovered by PTF, given PTF's observing epochs, conditions and the SN apparent brightness. Each artificial observation of a SN contributes a point on the light curve with a probability given by the single-epoch detection efficiencies. Light curve quality cuts can then be used to determine whether a simulated object would have met the requirements to enter our sample of real SNe Ia. This then gives us a second efficiency grid, this time indexed by the SN Ia intrinsic properties and redshift, from which we can calculate the efficiency factor, $\epsilon$, needed to re-weight the observed population in equation~\ref{eqn:rate}.

We discuss the details of this procedure over the next two sections. In section~\ref{sec:sn_sample_sec}, we discuss the PTF SN Ia sample and the measurements needed for our analysis, and in Section~\ref{sec:survey_sim} we discuss the details of our simulations.

\section{The PTF type Ia supernova Sample}
\label{sec:sn_sample_sec}

This section is concerned with the construction of the sample of PTF SNe Ia that enter our rate calculation. Constructing this sample requires us to overcome several challenges. The first is the SN classification: while many of our SNe are spectroscopically classified, a fraction were not observed spectroscopically due to a lack of follow-up resources, but must still be identified and included in our sample to calculate an accurate rate. The second challenge is to estimate, for each event, SN photometric parameters (such as light-curve shape) in order to accurately estimate the efficiency of PTF of recovering such an event. Accurate estimations of these parameters require data-quality cuts to ensure sufficient data is available. By applying the same cuts to both real and simulated data, we can ensure that our calculated efficiencies are directly applicable to our real dataset.

We begin by describing our definition of the photometric measurements of our SNe, and our light-curve fitting that determines the key photometric parameters. We then discuss the sample definition, and methods for including both spectroscopic and photometric events. We stress that while we describe the spectroscopic and photometric samples separately, all objects were considered against an identical selection function. Spectroscopic information, where available, is only to used to fix the object redshift when performing a light curve model fit.

\subsection{Photometric coverage cuts}
\label{sec:lc_fit}

Our goal is to replicate as closely as possible the pipeline that was used to initially discover PTF candidates, and on which our efficiency grid is based. Thus our coverage cuts are based on the real-time photometry produced by PTF, and ensure a high-quality sample of SNe that can both be reliably constrained by the light curve model and simulated in later analyses. Our coverage cuts are:
\begin{enumerate}
\item There must be at least four epochs on which the SN is detected. To be considered detected, each epoch must have an RB score ${\ge}0.07$, and $R<20$\,mag;
\item These four epochs must all be separated by $\ge12$ hours;
\item There must be at least two of these epochs prior to maximum brightness, and at least two of these epochs post maximum brightness.
\end{enumerate}
In addition, we require that for each of the four epochs, the median date of the reference image used in the real-time pipeline must be $\ge25$ days prior to the observation. The phases of each epoch were estimated using a light-curve fit to the data using the SALT2 SN Ia model, as implemented in the \textsc{python} package \textsc{sncosmo} \citep{barbary_2014_11938}.

Our single-epoch detection efficiencies are based on the real-time pipeline and measurement data, and hence the above coverage cuts are made on the same data. However, improved \lq forced-photometry\rq\ can also be measured, which typically significantly improves the quality of the photometry. This is particularly valuable for improving the light-curve (SALT2) model fits. Thus, for all events that pass our coverage cuts, we use the \textsc{ptfphot} pipeline, as widely used within the PTF community \citep[e.g.,][]{2011MNRAS.418..747M,2012MNRAS.426.2359M,2014MNRAS.438.1391P,2015MNRAS.446.3895F,2017MNRAS.468.3798D}, to produce higher-quality photometry. The pipeline is based on the same image subtraction principles as the real-time pipeline, but using a higher-quality deep reference image created from a larger combination of observations taken prior to the supernova explosion. The pipeline also uses PSF photometry (rather than \textsc{SExtractor} aperture photometry), with the PSF measured from isolated stars, and fits forced at the averaged SN position in all images. The photometry was then either calibrated to SDSS, or to the catalogues of \citet{2012PASP..124...62O}. We stress that all objects must pass the coverage cuts using their real-time photometry to enter our sample. Improving the photometry is only used to reduce the uncertainties in the SN Ia model parameters discussed in the following section.

\subsection{SN Ia population definition}

The SN Ia photometric parameters are related by 
\begin{equation}
  \label{eqn:lc_SALTeqn}
M_{B}=-19.05- \alpha x_1 + \beta \mathcal{C} + \sigma_\mathrm{int}.
\end{equation}
where M$_\mathrm{B}$ is the standardised absolute magnitude of an event, $-19.05$ is a typical SN Ia absolute magnitude, $x_1$ is a measure of the SN Ia light-curve width, $\mathcal{C}$ is the SN colour, $\sigma_\mathrm{int}$ is the intrinsic dispersion parameter that captures how a population of SNe Ia vary in brightness after standardisation corrections, and $\alpha$ and $\beta$ parametrize the SN Ia light-curve-width--luminosity and colour--luminosity relations.

We adopt the \citet{Betoule2014} definition of a \lq cosmologically useful\rq\ SN Ia, and thus we apply their selection criteria as closely as possible to the PTF objects. As a result, some objects may  be classified as a SN Ia based on features in their spectra, but photometric properties may exclude them from our analysis. We follow this selection criteria to construct a sample of `normal' SNe Ia, but do not intend to use them for an analysis of cosmological parameters.

We require $-3\le x_1\le3$ for all our SNe Ia, discarding SNe if they fall outside of this range. As PTF was principally a single band survey, we do not use colour information to define our sample. We also use a cut in absolute magnitude. As PTF measures light curve in the $R$-band, we construct an $M_R$ cut as follows. We use the SALT2 fits for the SNe Ia in \citet{Betoule2014}, and use them (along with nuisance parameters $\alpha=0.141$ and $\beta=3.101$) to construct a synthetic spectrum for each SN Ia at peak brightness. We then calculate $M_R$ for all the `joint light-curve analysis' (JLA) SNe Ia by integrating the synthetic spectrum through the $R_\mathrm{P48}$ transmission function. We performed ${>}$15,000 realisations of the data, drawing random parameters from the covariance matrix to construct a representative distribution of $M_R$. We find that $>$99 per cent of the JLA SN Ia population falls in the range $-19.75\le M_R\le -18.00$, and use this as our acceptable magnitude range for the PTF SN Ia sample.

\begin{figure}
	\centering
		\includegraphics[width=\linewidth]{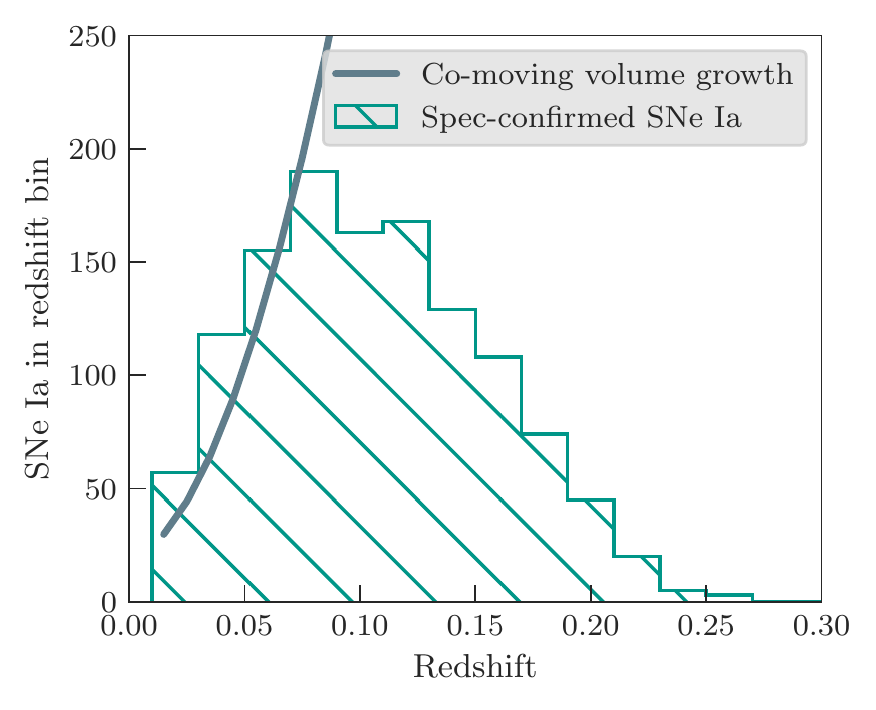}
	\caption{The redshift distribution of spectroscopically confirmed SNe Ia in PTF. The solid line traces a scaled co-moving element volume growth.}
	\label{fig:specDist}
\end{figure}

Finally, we set redshift limits for our sample. The lower limit is set as $z_1\ge0.015$, as this is the redshift at which a typical SN Ia would saturate the P48 detector (${\sim}$15\,mag). The upper-limit in redshift, depends on the level of spectroscopic incompleteness we can accept in our analysis. While the number of events increases with redshift, the spectroscopic completeness falls (and systematic uncertainty increases). Fig.~\ref{fig:specDist} shows the redshift distribution ($N(z)$) of spectroscopically confirmed SNe Ia in PTF, compared to the differential co-moving volume element of the universe. In a complete sample, we would expect the $N(z)$ to approximately scale following the co-moving volume element, which occurs until $=0.09$. The median redshift of all spectroscopically confirmed PTF SNe Ia is 0.1, and at $z=0.09$, a typical SN Ia has a peak apparent magnitude of $R\sim18.5$, $\simeq$2.5\,mag above the nominal PTF detection limit. This leads us to set $z_2=0.09$.

\begin{table}
\centering
\caption{The parameters and their acceptable ranges that define our SN Ia rate sample.}
\label{tab:rate_final_params}
\begin{tabular}{lc}
\hline
Parameter & Range        \\
\hline
$M_R$      &   $-19.75\leq M_R\leq-18.00$ \\
$x_1$      & $-3.00\leq x_1\leq3.00$         \\
$z$         & $0.015\leq z\leq0.090$  \\
\hline
\end{tabular}
\end{table}

Our final sample cuts are listed in Table~\ref{tab:rate_final_params}. We require that all PTF SNe Ia lie  within these limits.

\subsection{The SN sample}
\label{sec:sn-sample}

The follow-up resources available to PTF did not allow for every detected transient to be spectroscopically classified. We therefore categorise our SNe into two classes: those that were spectroscopically confirmed, and those that are photometrically typed. 

\subsubsection{The Spectroscopic Sample}
\label{sec:spec_sample}

SN spectra were taken on a number of facilities/instruments, predominantly the Palomar 200-in telescope with the Double Spectrograph \citep{1982PASP...94..586O}, the Keck 10-m telescope with the Low-Resolution Imaging Spectrometer (LRIS) \citep{1995PASP..107..375O}, the 4.2-m William Herschel Telescope with the IRIS spectrograph, and the Lick Observatory Shane 3-m telescope with the Kast double spectrograph \citep{miller_lick_1993}. PTF SN Ia spectra are presented in \citet{2014MNRAS.444.3258M} and available on WISeREP\footnote{https://wiserep.weizmann.ac.il} \citep{2012PASP..124..668Y}. Across all redshifts, a total of 1,241 SNe Ia were spectroscopically classified, and entered our preliminary sample.

\subsubsection{Photometric SNe Ia}
\label{sec:phot_search}

We next search for potential SNe Ia in the objects for which PTF did not obtain a spectrum. This requires a search of the entire PTF transient database\footnote{Our search was performed on an internal PTF database; however, the entire source and light-curve tables are publicly available from \url{http://irsa.ipac.caltech.edu/}}, a database of $\sim$48,000 unique candidate objects (without spectra). This dataset is not  populated solely by SNe, but also by spurious detections, variable stars, and active galactic nuclei. Our goal is to identify objects that are likely SNe Ia and that should enter our rate sample. The cuts applied to the data are summarised in Table~\ref{tab:phot_cand_cuts}.

We begin by applying similar coverage cuts as described in Section~\ref{sec:lc_fit} to all objects in the database: each event detected with $m_R<20$ on at least 4 epochs each separated by at least 12 hours. This reduced the number of candidates to ${\sim}$5,000. We then perform SALT2 model fits to each event, but in the absence of redshift information we leave this parameter free and apply a uniform prior between $0<\mathrm{z}<0.3$ on the fit, as a very conservative upper limit for SNe Ia in PTF.

The vast majority of the fits were very poor, as most of the objects were spurious detections at low signal-to-noise. We thus make a very conservative cut on the goodness-of-fit of $\chi^2_\nu\le150$, reducing to 1,117 objects. We applied our final coverage cut on each light curve, where each object must have at least 2 epochs on the rise and 2 epochs on the decline of the light curve within the valid phase range of the SALT2 model. This further reduced our sample to 216 objects.

We then visually inspected the light curves of the candidates, and used contextual information such as a pre-existing variable source at the candidate location. This contextual assessment came from the PTF reference images and image cut-outs from the Sloan Digital Sky Survey \citep[SDSS;][]{2000AJ....120.1579Y,2011AJ....142...72E}. Furthermore, our candidates were compared against galaxies with spectroscopic redshifts in the SDSS database, and visually crossed-checked in the event of host mis-dentification \citep{2010MNRAS.406..782S,2016AJ....152..154G}. We found 59 of our candidate SNe in galaxies with a spectroscopic redshift, but only 7 in our redshifts within our boundary defined in Table~\ref{tab:rate_final_params}; thus we remove 52 of the unknown candidates from our analysis. Our penultimate cut removed all candidates that did not appear in our 8 areas-of-interest, described in Section~\ref{sec:survey_area}, and in total we were left with 31 candidate SNe Ia.

Our final cut is to ensure that the SALT2 fit parameters of all the events lie in the ranges in Table~\ref{tab:rate_final_params}, holding the redshift fixed at the spectroscopic redshift where available. Only 6 of the photometric objects pass the cuts we define for normal SNe Ia, and these are presented in Table~\ref{tab:best_ptfname_photometric}. Our light curve cuts enforce a theoretical lower limit of 9 days of coverage. The shortest duration, however, between the first and final detection of a SN in our sample is 17 days.

\begin{table}
\centering
\caption{The cuts made to identify potential photometric SNe Ia candidates.}
\label{tab:phot_cand_cuts}
\begin{tabular}{lr}
\hline
Cut                         & Objects remaining \\\hline
Photometric Database        & 48,474      \\
4 epochs with $m_R<20$      & 5,051             \\
SALT2 Goodness-of-fit       & 1,117             \\
2 epochs on rise, 2 on decline & 216               \\
Host spec-z $\le 0.09$        & 164               \\
In area-of-interest         & 31                \\
SALT2 fit parameters        & 6                 \\
\hline
\end{tabular}
\end{table}

\begin{table*}
\centering
\caption{Photometrically identified SNe Ia in the PTF rates sample.}
\label{tab:best_ptfname_photometric}
\begin{tabular}{lrrrrrrc}
\hline
PTF Name & R.A. (J2000)     & Decl. (J2000)     & z     & t$_0$ (MJD)         & x$_1$ & M$_R$  & Host spec-z \\\hline
PTF10lrp    & 14:51:23.42 & +35:45:39.42  & 0.079 & 55375.47 & -0.18 & -19.13 & True                 \\
PTF12gcn    & 15:23:43.73 & +08:26:14.09  & 0.075 & 56111.91 & -0.23 & -18.48 & True                 \\
PTF10gxa    & 16:34:28.30 & +57:36:24.05 & 0.064 & 55321.18 & 1.10  & -18.27 & False                \\
PTF11cxe    & 16:43:15.82 & +40:32:09.31 & 0.066 & 55693.39 & -0.79 & -18.06 & False                \\
PTF12dah    & 15:21:48.91 & +50:04:21.43 & 0.084 & 56043.48 & 1.81  & -18.74 & False                \\
PTF12cci    & 14:15:29.30 & +39:54:08.93  & 0.067 & 56024.95 & 0.86  & -19.06 & True \\
\hline
\end{tabular}
\end{table*}

\subsection{Local surface brightness}
\label{sec:sam_surface_brightness}

We also measure the local host-galaxy surface brightness at the position of all PTF SNe Ia. We use the parameter \fboxsb\ as defined by \citetalias{2017ApJS..230....4F}: this measures the flux in a $3\times3$ pixel box (9.18$\arcsec^2$) at the location of the SN before the SN exploded. \citetalias{2017ApJS..230....4F} showed that the detectability of transients was affected by the immediate surface brightness, with the faintest objects less often recovered when occurring in the brightest environments. It is, therefore, a necessary parameter for us to consider when calculating our recovery efficiencies for SNe in our sample. We make these measurements on the deep reference images generated by the \textsc{ptfphot} pipeline. \fboxsb\ is defined as
\begin{equation}
\label{eqn:fbox_to_sb_eqn}
\fboxsb =  -2.5\log_{10}\left(\frac{\flbox}{9.18}\right) + \mathrm{zp}
\end{equation}
where \flbox\ is the sum of the counts in the $3\times3$ pixel box, and zp is the zeropoint of the image.

\section{Survey Simulation}
\label{sec:survey_sim}

We now describe our simulations to determine how well PTF recovers SNe Ia as a function of the SN Ia light curve properties and local surface brightness. We use Monte Carlo simulations of the survey operation, with artificial SN light curves generated and their recovery rate measured. Events from the observed PTF SN Ia population can then be compared to these artificial SNe, allowing an efficiency to be assigned to each PTF SN Ia event.

The artificial SN Ia light curves are simulated using the real observing conditions of PTF, and then using the work of \citetalias{2017ApJS..230....4F} to statistically accept or reject points on a simulated light curve.  

\subsection{Survey Area and Duration}
\label{sec:survey_area}

\begin{figure*}
	\centering
		\includegraphics[width=\linewidth]{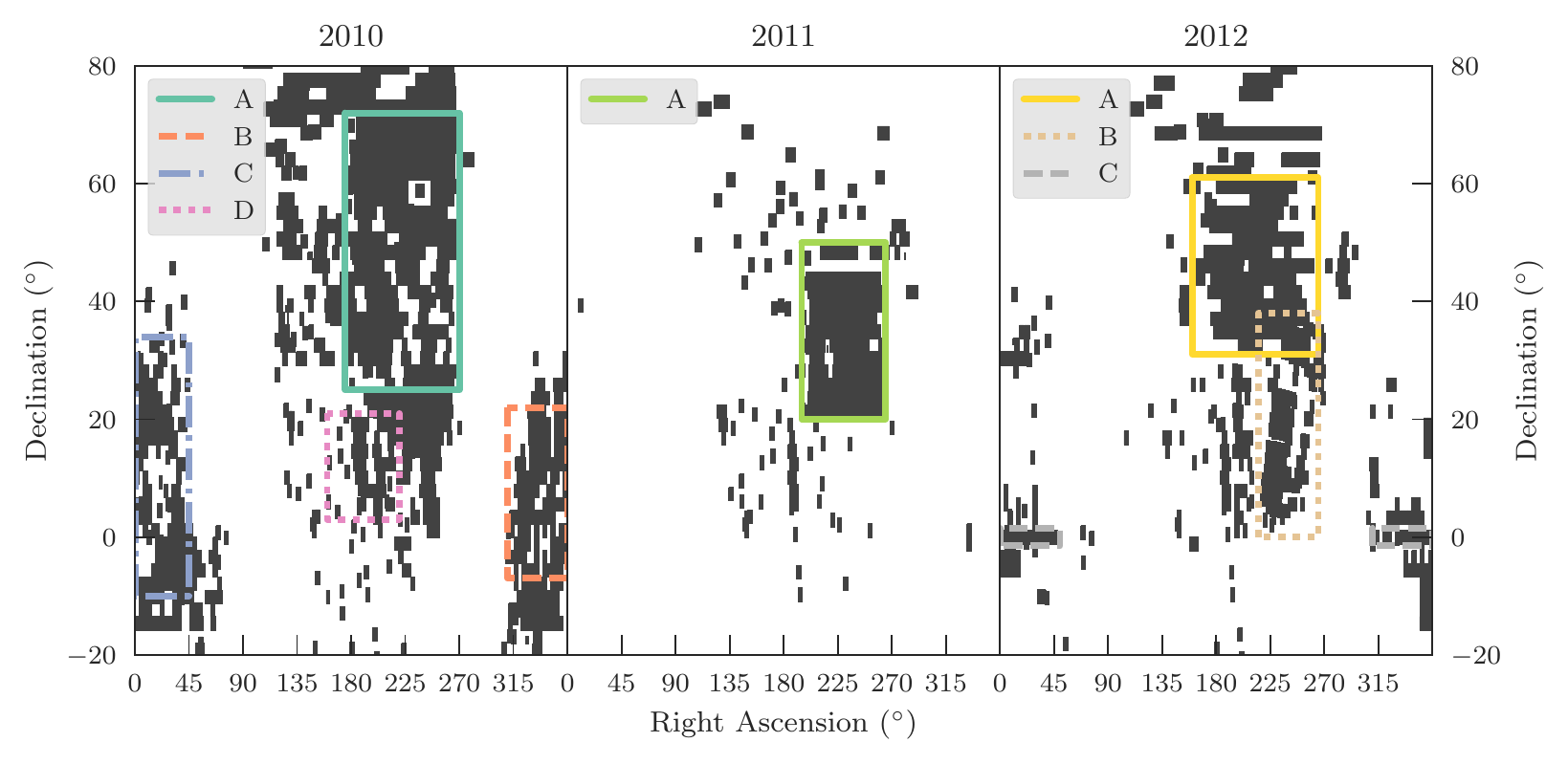}
	\caption{The survey area of PTF in 2010, 2011, and 2012. Each dark tile represents a field observed at least once during each season. The boxes show the footprints in which we choose to simulate SNe Ia in this rates analysis. These footprints were designed to bound areas of regular cadence for extended periods.}
	\label{fig:skySimBoxes}
\end{figure*}

We first define the survey area and duration. While previous rates analyses, such as  \citet{2008ApJ...682..262D,2010ApJ...713.1026D,2012AJ....144...59P,2011MNRAS.412.1419L,2014ApJ...783...28G}, were performed using surveys with a well defined cadence and/or a fixed area of observation, the PTF search fields evolved over the course of the survey over thousands of square degrees. We established 8 areas on the sky, across 2010--2012 in the 5 day cadence seasons, where PTF maintained both a regular cadence and a consistent observing footprint (see Table~\ref{tab:lc_sky_area_nights} for details, and a visualisation in Fig.~\ref{fig:skySimBoxes}). The choice of these areas is clearly somewhat arbitrary; however, ultimately this only impacts the efficient use of computational resources for the simulations. The data in Table~\ref{tab:lc_sky_area_nights} can then be used to construct the volumes and $\Delta T$ required in equation~\ref{eqn:rate}.

\begin{table}
\centering
\caption{The simulation areas and durations in this rates analysis (Fig.~\ref{fig:skySimBoxes}).}
\label{tab:lc_sky_area_nights}
\begin{tabular}{clrr}
\hline
& Box    & Area & Duration\\
& &(deg$^2$)&(days)\\
\hline
\textcolor{2010A}{$\blacksquare$} & 2010 A & 2521.81 & 170             \\
\textcolor{2010B}{$\blacksquare$} & 2010 B & 1248.79 & 140             \\
\textcolor{2010C}{$\blacksquare$} & 2010 C & 1680.52 & 109             \\
\textcolor{2010D}{$\blacksquare$} & 2010 D & 611.37  & 149             \\
\textcolor{2011A}{$\blacksquare$} & 2011 A & 1495.10 & 150             \\
\textcolor{2012A}{$\blacksquare$} & 2012 A & 1664.89 & 109             \\
\textcolor{2012B}{$\blacksquare$} & 2012 B & 918.45  & 78              \\
\textcolor{2012C}{$\blacksquare$} & 2012 C & 299.97  & 139      \\      
\hline
\end{tabular}
\end{table}

\subsection{Light Curve Simulations}
\label{sec:lc_sim}

The method of simulating SN Ia light curves in PTF data is described in \citetalias{2017ApJS..230....4F}.
In summary, SNe Ia are generated following the model of equation~\ref{eqn:lc_SALTeqn}, drawing random values for $x_1$ uniformly and $\mathcal{C}$ from the population distribution in \citet{Betoule2014}, and assigning a random redshift ($z$), R.A., Decl., and explosion date to each event. A value for the intrinsic-dispersion, $\sigma_\mathrm{int}$, is randomly drawn from a normal distribution with $\mu=0$ and $\sigma=0.15$. These parameters then define a synthetic spectral time-series that we convert, with appropriate K-corrections, to an apparent magnitude in the $R_{\mathrm{P48}}$ band. Based on the SN position, we apply Milky Way extinction \citep{1998ApJ...500..525S} following \citet{1989ApJ...345..245C}. We then query the PTF observing history to determine the epochs (and observing conditions) on which it would have been observed.

The \citetalias{2017ApJS..230....4F} single-epoch recovery efficiencies are then used to assign a detection probability on each epoch, and the resulting synthetic light curve checked to determine whether it passes the light-curve quality cuts in Section~\ref{sec:lc_fit}. This method of simulating the recovery of light curves is fast as we do not insert the light curves into the images. Instead, we use the statistical properties of the single-epoch recovery efficiencies, and the observing meta-data, to make a probabilistic statement about the simulated object's detection. By generating many realisations of the light curve variables and redshift, the SN Ia recovery fraction as a function of the light-curve properties can be determined. We extend this technique to include a consideration of the local surface brightness parameter, \fboxsb. We perform the light-curve simulation as described above, but with 7 different values for \fboxsb. This allows us to compare the recovery fraction of SNe Ia as a function of host brightness. In total, we simulated $>4.6{\times}10^8$ artificial SNe Ia, with each analysed under 7 different host surface-brightness assumptions. 

\subsection{Efficiency Grids}
\label{sec:rate_efficiency}

We use these simulation results to define a multi-dimensional grid that describes the fraction of simulated events recovered, as a function of several of the simulated parameters. Each real SN Ia in our PTF sample occupies a location within each grid, providing an (interpolated) efficiency, $\epsilon$, for each object.

\begin{figure*}
	\centering
		\includegraphics[width=\linewidth]{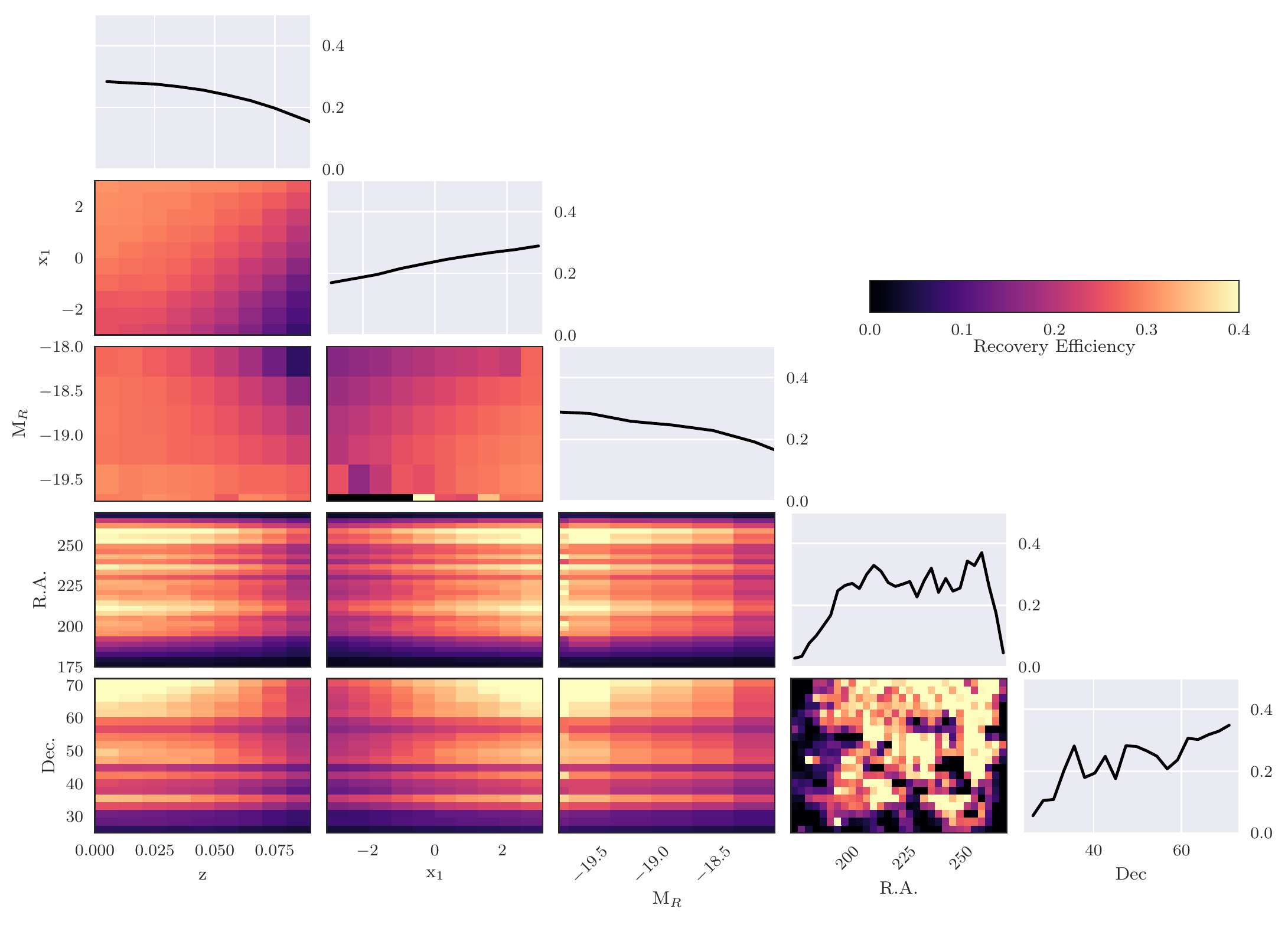}
	\caption[The multidimensional efficiency grid for 2010 A]{The final efficiency grid for the 2010A field and the faintest \fboxsb. The diagonals are the 1D efficiencies for each of the grid parameters, and the off diagonals are efficiencies for combinations of those parameters.}
	\label{fig:lc_eff_grid_2010A}
\end{figure*}

Our SNe Ia are defined in terms of $x_1$, $M_R$, and $z$, and these form the first three axes of our efficiency grid. Our 8 simulation areas (Fig.~\ref{fig:lc_eff_grid_2010A}) contain unobserved patches, which leads to a non-uniform spatial recovery efficiency; we therefore include an R.A. and Decl. dependence in the efficiency grids. By design, our simulation areas were observed with a consistent cadence, this minimises time-dependent variations in the recovery efficiency, so we do not consider time a parameter in our grids. The final variable is the host-galaxy surface brightness, and we construct a different efficiency grid for each of our 7 \fboxsb\ values. This gives 56 different grids that parametrize the PTF survey for SNe Ia.  The efficiencies we determine (for $x_1$, $M_R$, $z$, R.A. and Decl.) from our grids are insensitive to the underlying parameter distributions. We are unable to measure colour for our real SNe, but use a realistic underlying distribution for $\mathcal{C}$  \citep{Betoule2014} in our simulations and marginalise over its effect.

In Fig.~\ref{fig:lc_eff_grid_2010A}, we show an example efficiency grid constructed from the faintest \fboxsb\ bin. We see the expected trends: SNe Ia at higher redshifts show a decreasing recovery efficiency due to their decreasing apparent brightness; SNe with a larger $x_1$ value are recovered more frequently; and SNe Ia with a brighter $M_R$ are recovered with a higher efficiency. The final parameters on the grid, R.A. and Decl., show the importance of considering the spatial variation. The alternative -- ignoring the R.A. and Decl. variation -- would be equivalent to marginalising over these dimensions, averaging the spatial efficiencies and systematically lowering the recover efficiency for the entire field. 

\begin{figure*}
	\centering
		\includegraphics[width=\linewidth]{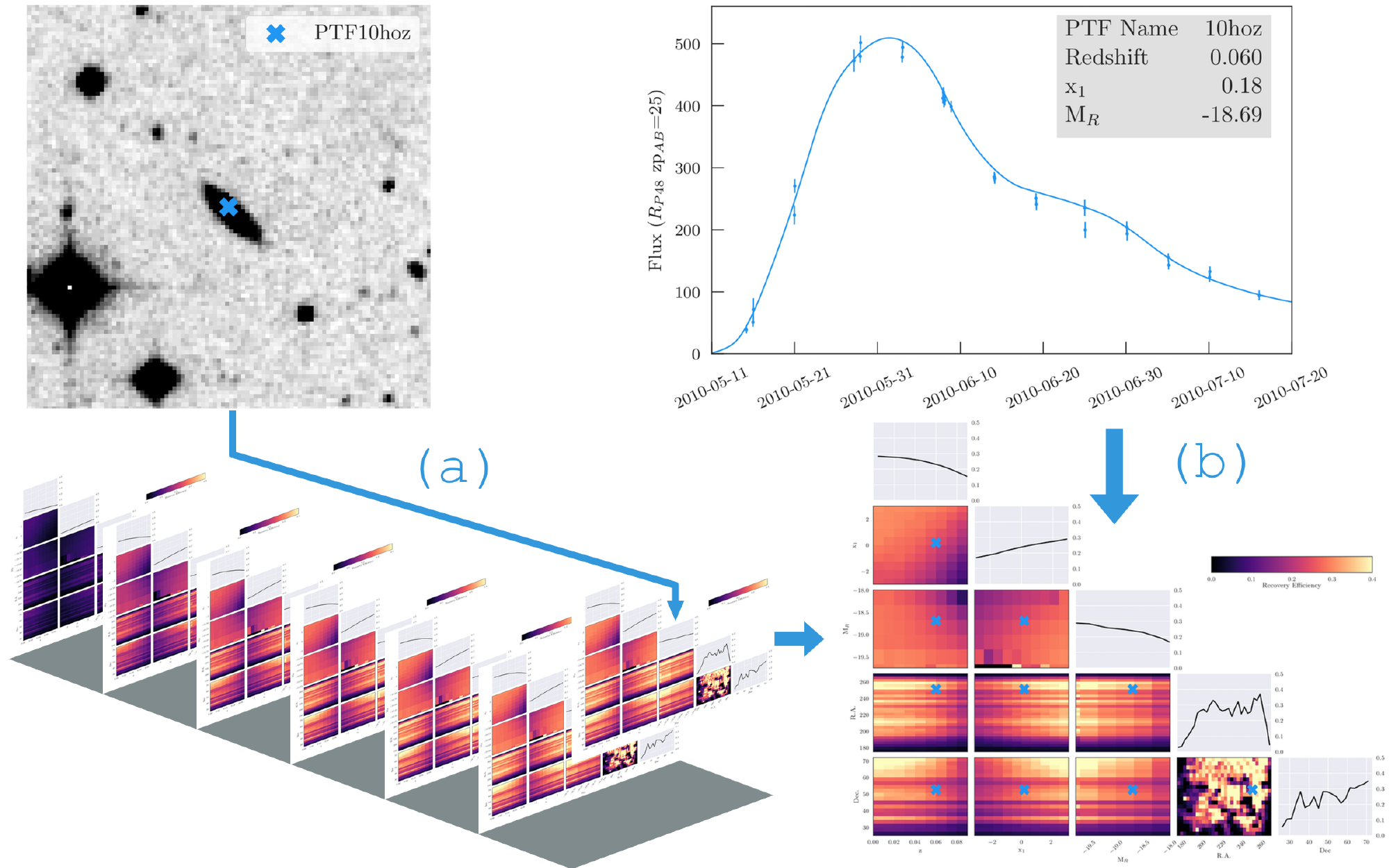}
	\caption{A schematic of the technique used to determine the object efficiency weighting for an example PTF event, PTF10hoz. In stage (a) the location and immediate surface brightness, \fboxsb, from the reference image is used to select the appropriate efficiency grid. In stage (b), the real values of $z$, $x_1$, $M_R$, R.A., and Decl. are used along with a multi-dimensional linear interpolation to calculate the efficiency parameter, $\epsilon$. This schematic uses the efficiency grid from Fig.~\ref{fig:lc_eff_grid_2010A}.}
	\label{fig:rate_eff_diagram}
\end{figure*}

The method for combining these efficiency grids for our real SNe is schematically presented in Fig.~\ref{fig:rate_eff_diagram} for one event. We use the set of efficiency grids (each for a different \fboxsb) generated for each area, choose the grid that contains the observed \fboxsb\ parameter, and perform a multi-dimensional linear interpolation at the real SN $z$, $x_1$, $M_R$, R.A. and Decl. to calculate the recovery fraction, $\epsilon$. This was repeated for all SNe Ia in the sample.

\begin{figure*}
	\centering
		\includegraphics[width=\linewidth]{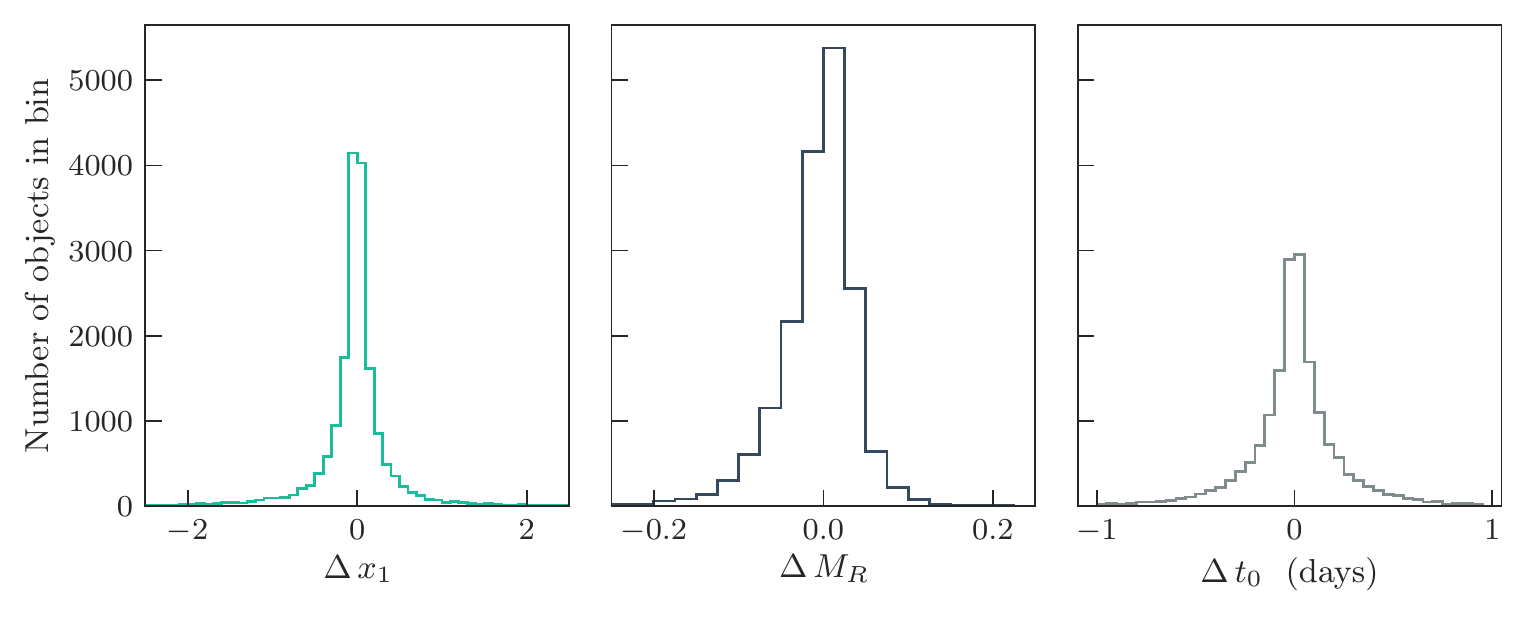}
	\caption[Comparing Sim to Recovered]{The difference between the input simulation parameters and the SALT2 fits for the `observed' simulated light curves are shown for our measurable parameters $x_1$, $M_R$, and $t_0$. Our investigation was performed on a sub-sample of ${\sim}$18,000 simulated objects that met our light curve quality cuts. These objects were fit with the SALT2 model and compared to the original parameters from which they were generated. All three distributions peak at zero difference and confirm there is no systematic bias introduced by using our simualted parameters to construct efficiency grids.}
	\label{fig:Sim2Recovered}
\end{figure*}

We now investigate any systematic effects introduced by constructing the efficiency grids with simulation parameters rather than the observationally derived results.  We performed a simulation of $\approx$\,18,000 light curves following the methods detailed in Section~\ref{sec:lc_sim}. To each of these objects we applied realistic observational uncertainties. The SNe were then fit with the SALT2 model, as described in Section~\ref{sec:sn_sample_sec}, with $z$ fixed at the simulation redshift to replicate information from a spectroscopically confirmed object. In Figure~\ref{fig:Sim2Recovered} we show how the fit-parameters differ from the `ground-truth' and find all our measurable parameters are peaked around zero. This suggests there is no large systematic bias, or shift, in the SN recovery efficiencies we derive from constructing efficiency grids with simulation parameters. This is an encouraging result for our methodology, as fitting all $4.6\times10^{8}$ artificial SNe with SALT2, and constructing grids from those measured parameters, would require a prohibitive amount of computational resource.

\section{The Volumetric Supernova Type Ia Rate}
\label{sec:rate_rate}

We now use the methodology and SN Ia sample defined in the previous sections to calculate the volumetric SN Ia rate, including an estimation of the rate uncertainties from a Monte Carlo simulation. In total, our sample consists of 90 SNe Ia that pass all cuts, 84 of which were spectroscopically identified, and 6 which were photometrically identified. The volume-weighted mean redshift of the sample is $z=0.073$. Using equation~\ref{eqn:rate} and the efficiencies as calculated in Section~\ref{sec:rate_efficiency} we calculate the volumetric rate of SNe Ia in PTF to be $r_V=2.42\times10^{-5}\,\mathrm{SNe}\,\mathrm{yr}^{-1}\,\mathrm{Mpc}^{-3}\, h_{70}^{3}$.

With 90 events and each SN carrying equal weight, we could approximate the statistical uncertainty as $\sqrt(N)$, where $N$ is the number of events. However, as our efficiencies vary from SN to SN, the weight of each event is not equal. We set the weight of each SN in the uncertainty calculation to $\omega=1/\epsilon$,  normalise the sum of the weights to 1, and calculate the statistical uncertainty $\sigma_{\mathrm{stat}}$ as
\begin{equation}
\label{eqn:rate_stat_err}
\sigma_{\mathrm{stat}} = \sqrt[]{\sum \omega_i^2} \times N.
\end{equation}
Our rate measurement is then
\begin{equation*}
r_v=2.42\pm0.29\times10^{-5}\,\mathrm{SNe}\,\mathrm{yr}^{-1}\,\mathrm{Mpc}^{-3},
\end{equation*}
at a volume-weighted redshift of $z=0.073$. These uncertainties are statistical only, and we next estimate our systematic uncertainties.

\subsection{Systematic Uncertainties}
\label{sec:sys_err}

Our rate was calculated using efficiencies derived for the best-fitting SALT2 values and redshift for each object in the SN Ia sample. The SALT2 fit parameters also carry uncertainty, and we account for this by performing a Monte Carlo simulation using the SALT2 fit covariance matrix to draw realisations of our light curves for each object. This can be understood by visualising each SN Ia as occupying a region of the efficiency grid, with a multi-dimensional probability distribution of efficiencies, $\epsilon$, described by the SALT2 model. Our Monte Carlo simulation draws efficiencies according to this distribution, and propagates them through the rate calculation, with a new volumetric rate calculated each time. This builds up a distribution of rate results that includes the light-curve fit parameter uncertainty. Furthermore, SNe with fit parameters near the boundary of the allowed ranges were allowed to enter or leave the SN Ia sample for each new realisation of the light curve parameters, perhaps changing the final sample size on each realisation. (Note that we always define the statistical uncertainty as that calculated from the best-fitting results.)

\begin{figure*}
	\centering
		\includegraphics[width=\linewidth]{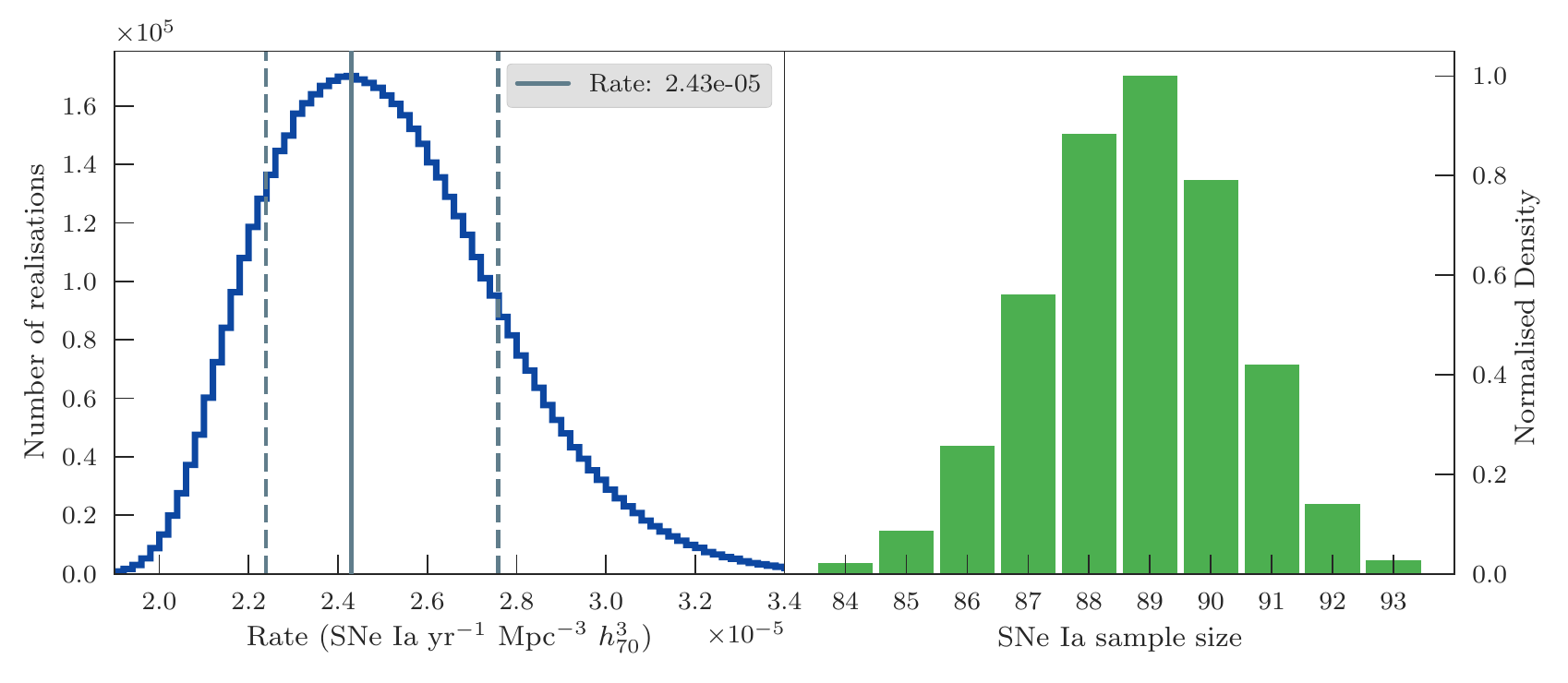}
	\caption{Left: The distribution of the SN Ia rate from our Monte Carlo Simulation of the SN sample. We use the distribution to quantify our systematic uncertainties due to model variation in the light curve fitting process. The solid vertical line represents the rate, and the two dashed vertical lines bound our 1-$\sigma$ confidence on the result. Right: The distribution of the SN sample size for the Monte Carlo simulation. The change in the sample size is caused by light curve realisations no longer meeting our criteria for a cosmically-useful SN Ia. The best-fitting light curve models produce a sample size of 90 objects, from which we calculate our statistical uncertainties, whereas the Monte Carlo simulation suggests a sample of 89 objects. This small sample size difference does not affect the statistical uncertainties in our calculation.}
	\label{fig:rateDistribution}
\end{figure*}

In total, $5.6\times10^6$ realisations of the rate were calculated, and the distribution shown in Fig.~\ref{fig:rateDistribution}. The final value of the rate is taken as the peak of the distribution and the uncertainties capture 68.3 per cent of the probability around the mean. The volumetric rate is then
\begin{equation*}
r_v=2.43\pm0.29\,\text{(stat)}_{-0.19}^{+0.33}\text{(sys)}\times10^{-5}\,\text{SNe\,yr}^{-1}\,\text{Mpc}^{-3}\, h_{70}^{3}.
\end{equation*}
The range in the number of SNe entering the rate from realisation to realisation is rather narrow (Fig.~\ref{fig:rateDistribution}), with a mode of 89 objects.

\section{Analysis}
\label{sec:analysis}

\subsection{Comparison to other rates}
\label{sec:rates_compare}

We first compare our PTF SN Ia rate to other survey measurements. We chose to only include rates with similar definitions of SNe Ia, and calculated using a comparable methodology to ours. This necessarily excludes the low-redshift rates of \citet{1999A&A...351..459C}\footnote{$r_v =2.8\left(\pm0.9\right)\times10^{-5}\,\text{SNe Ia\,yr}^{-1}\,\text{Mpc}^{-3}$, z=0} and \citet{2011MNRAS.412.1473L}\footnote{$r_v=3.01\left(\pm0.61\right)\times10^{-5}\,\text{SNe Ia\,yr}^{-1}\,\text{Mpc}^{-3}$, z=0} for; (i) being targeted surveys, and (ii) for contaminants by non-cosmically useful SN Ia sub-types, e.g. 91bg-like. Low-z SN surveys typically operate a galaxy targeted search for supernovae, with a bias towards observing the brighter, more massive galaxies. This potentially produces systematics in the observed SN population as SN Ia light curves and host properties are correlated \citep{2010MNRAS.406..782S}. The resulting SN rates are converted from functions of the host-galaxy properties to volumetric rates using a galaxy luminosity function. This procedure may introduce a preference for particular SNe Ia and bias the rate result. PTF operated in a untargeted rolling search mode and, therefore, does not suffer from this observational bias.

We additionally exclude the SDSS DR7 rate of \citet{2013MNRAS.430.1746G}. Although this was calculated from a novel search for contaminant SN flux in galaxy spectra, it is by definition galaxy-targeted, and there are no SN light curves. We therefore cannot check our agreement between survey sample definitions. We note however, that their result\footnote{$r_v=2.47_{-0.26}^{+0.29}\text{(stat)}_{-0.3}^{+0.16}\text{(sys)}\times10^{-5}\,\text{SNe Ia\,yr}^{-1}\,\text{Mpc}^{-3}$, z${\sim}$0.1} is in good agreement with our rate.

We now turn our attention to several comparable rates analyses from untargeted surveys and with a similar SN selection function to our own. The Supernova Legacy Survey (SNLS), a five-year rolling high-redshift SN search, conducted a rates analysis using 691 events \citep{2012AJ....144...59P} over $0.2\le z\le 1.1$, and remains the best measurement of the rate at these redshifts. The SN Ia rates of \citet{2011MNRAS.417..916G} were a reanalysis of \citet{2007MNRAS.382.1169P} from the Subaru Deep Field \citep{2004PASJ...56.1011K}. This high-redshift search found 28 SNe Ia in the redshift range $z=1-1.5$, and 10 over $z=1.5-2.0$. The rate of SNe Ia beyond $z=2$ was presented by \citet{2014AJ....148...13R} using data from the Cosmic Assembly Near-infrared Deep Extragalactic Legacy Survey (CANDELS) \citep{2011ApJS..197...35G,2011ApJS..197...36K}. Their sample of SNe Ia consisted of $\approx24$ objects, with the highest redshift bin coinciding with the peak of the cosmic star-formation history (SFH). 

Most comparable to our PTF rate, is the work of \citet{2010ApJ...713.1026D}, from the SDSS-II Supernova Survey \citep{2008AJ....135..338F}. This sample contained 516 SNe Ia, of which 270 were spectroscopically confirmed out to a redshift of $z\sim0.3$. The lowest redshift bin in their study contained only four objects, while the bin closest in redshift to our PTF result contained 31 SNe Ia. Their statistical uncertainties dominated the low-redshift error budget ($z\le0.2$). The larger sample of PTF SNe (90 objects) provides improvements on the Poisson uncertainty, and our method of weighting individual objects contrasts with the SDSS measurement which bulk corrected the SN sample based on simulations using light curve parameter distributions. This differing treatment of the detection efficiencies ultimately produced a consistent result of the local SNe Ia rate.

\begin{figure}
	\centering
		\includegraphics[width=\linewidth]{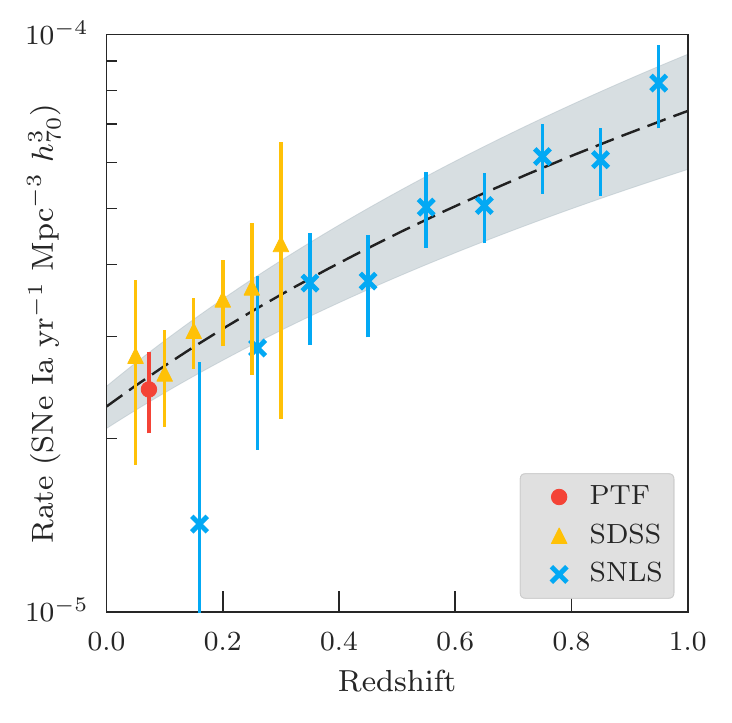}
	\caption{The PTF rate (circle) compared to other, non-targeted, SN survey results, SDSS \citep[triangles;][]{2010ApJ...713.1026D} and SNLS \citep[crosses;][]{2012AJ....144...59P}. The dashed line shows the power-law fit to the data of $r_v(z) = r_{0}(1+z)^{\alpha}$. We find $r_0 = 2.27\pm0.19\times 10^{-5}$ and $\alpha=1.70\pm0.21$. The grey shaded region represents the 1-$\sigma$ uncertainty on the rate evolution. We discard the $z>1$ SNLS data following \citet{2012AJ....144...59P}.}
	\label{fig:ratepowerlaw}
\end{figure}

Our final compilation of literature rates comprises of; SDSS \citep{2010ApJ...713.1026D}, SNLS \citep{2012AJ....144...59P}, SDF \citep{2011MNRAS.417..916G}, and CANDELS \citep{2014AJ....148...13R}. We adjust these published rates to the cosmology assumed here, and fit a simple power-law to the redshift evolution of the rate for $z<1$,  modelled as
\begin{equation}
\label{eqn:rate_evolution}
r_v(z)=r_{0}(1+z)^{\alpha}.
\end{equation}
We find $r_0 = 2.27\pm0.19\times 10^{-5}$ and $\alpha=1.70\pm0.21$. This result is consistent with the similar analysis of \citet{2012AJ....144...59P}. 

\subsection{The Delay-Time Distribution}
\label{sec:DTD}
In this final section, we briefly investigate the implications of our improved low-redshift measurement in the context of the SN Ia delay-time distribution (DTD). The SN Ia DTD describes the SN Ia rate as a function of time following an instantaneous burst of star formation. SNe Ia are the impulse response to star formation, and the time from progenitor birth to explosion is determined by the underlying progenitor physics. Thus, in the absence of a direct observation of a SN Ia system pre-explosion, SN Ia rates can provide an alternative  insight into possible progenitor systems. The SN Ia rate as a function of cosmic time ($\mathrm{SNR}_\mathrm{Ia}(t)$) can be modelled as the convolution of the DTD ($\Psi$) and the SFH, i.e.,
\begin{equation}
\label{eqn:rate_dtd}
\mathrm{SNR}_\mathrm{Ia}(t) = \int_{0}^{t} \mathrm{SFH}(t-\tau)\Psi(\tau) d\tau
\end{equation}
where $\tau$ is the delay-time, and t is the elapsed time from progenitor-star birth.

Different progenitor scenarios are expected to produce different DTDs \citep[see][for a review]{2012PASA...29..447M}. A popular parametrization is a simple power-law, $\Psi (t)=t^{-\beta}$, and we fit this form here. We set the first 40\,Myr of the DTD to zero, as this corresponds to the expected life-time of stars with an initial mass greater than 8\,M$\odot$ (expected to explode as core collapse SNe). Systematic uncertainties on the DTD are dominated by the SFH \citep{2006MNRAS.368.1893F,2011MNRAS.417..916G,2014ApJ...783...28G}, and so we explore three different SFH parametrizations. We use the \citet{2008MNRAS.388.1487L} SFH in both the piecewise form and the \citet{2001MNRAS.326..255C} functional form, and the \citet{2008ApJ...683L...5Y} SFH. We then fit to our rate compilation, and the results for different assumed SFHs are shown in Table~\ref{tab:rate_dtd_best_fit}.

\begin{table}
\caption{Best-fitting t$^{-\beta}$ values for the DTD convolved with different assumptions of the SFH.}
\label{tab:rate_dtd_best_fit}
\begin{tabular}{l|c}
\hline
Star Formation History Model               & $\beta$       \\
\hline
\citet{2008MNRAS.388.1487L} (piecewise) & $0.95\pm0.07$ \\
\citet{2008MNRAS.388.1487L} + \citet{2001MNRAS.326..255C} & $1.07\pm0.1$  \\
\citet{2008ApJ...683L...5Y}         & $0.94\pm0.08$\\
\hline
\end{tabular}
\end{table}

Clearly, the power-law description of the DTDs are all consistent with $\beta=1$, regardless of our SFH choice. Such a power-law is consistent with a double degenerate (DD) progenitor population, as also found by many similar studies \citep{2010ApJ...713.1026D,2012AJ....144...59P,2013MNRAS.430.1746G,2012MNRAS.426.3282M,2014ApJ...783...28G}, although this does not exclude all single-degenerate progenitor models \citep[e.g.][]{2008ApJ...683L.127H}.

\begin{figure*}
	\centering
		\includegraphics[width=\linewidth]{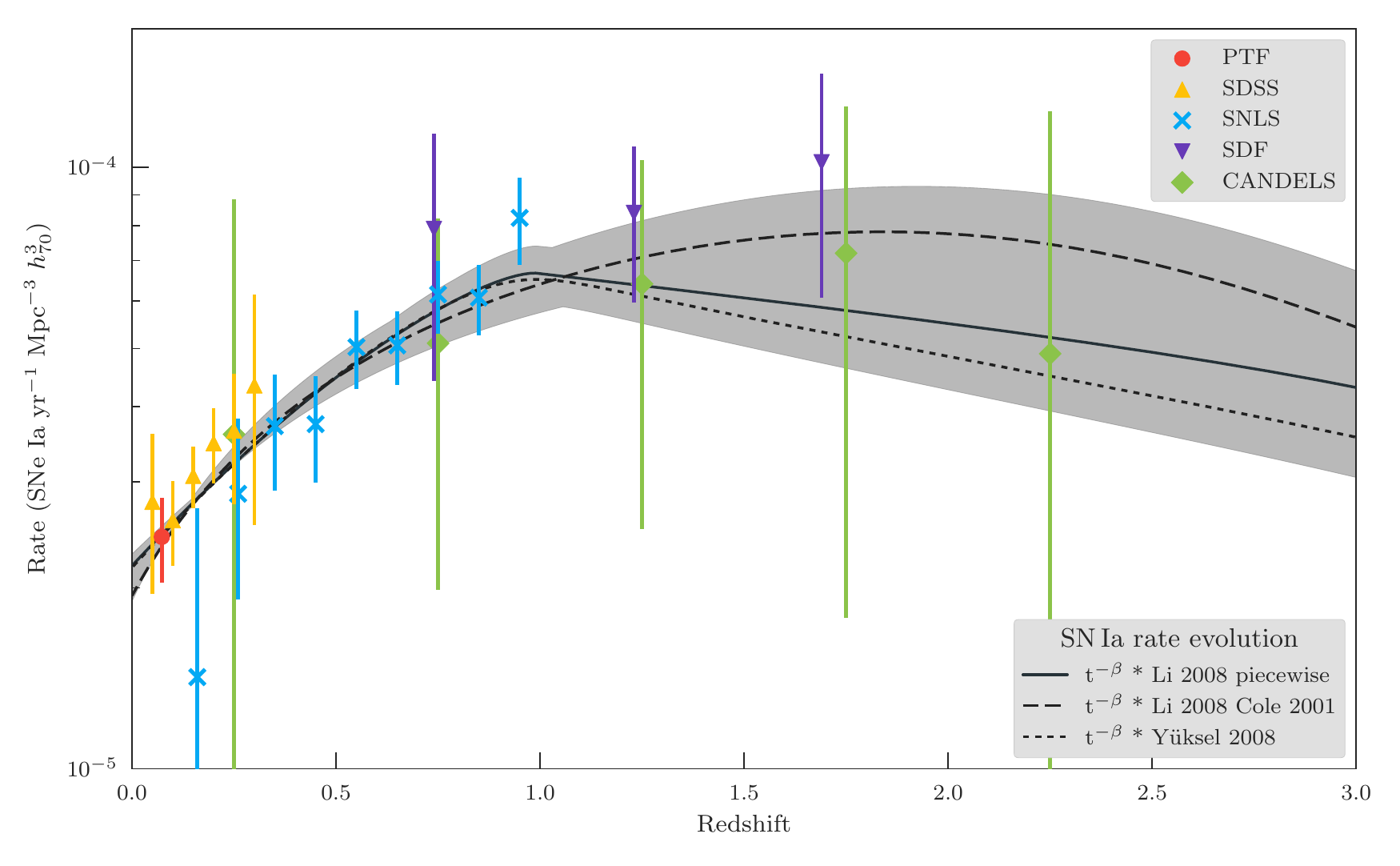}
	\caption{The best-fitting rate evolution for a simple power-law $t^{-\beta}$ DTD. All values of $\beta$ were consistent with 1. The grey regions show the combined 1-$\sigma$ uncertainties on the fit resulting from the different SFH models. The data are well constrained at low redshift, but improved rate measurements at $z>1$ would help constrain the SN Ia rate turnover.}
	\label{fig:rate_dtd_fit}
\end{figure*}

The PTF SN rate adds a well-constrained and well-understood low-redshift measurement for the rate evolution of normal SNe Ia, independent of other survey results that operated over a wide redshift range. Further improvements to our understanding of SN Ia progenitors would come from equally well-constrained high-redshift ($z>1$) rate measurements. This may help motivate new DTD models that could explain whether different SN Ia progenitors operate at different redshifts. 

\section{Summary}
\label{sec:summary}

This paper has presented the volumetric type Ia supernova (SN Ia) rate in the local universe ($z\le0.09$) using data from the Palomar Transient Factory (PTF). Our method of calculation builds on the detection efficiency study of \citet{2017ApJS..230....4F} to create a statistically representative simulation of the PTF survey using a Monte Carlo approach. We simulated the observations for several hundred million SNe Ia to quantify their detectability as a function of redshift, light-curve parameters, and host galaxy surface brightness. Our simulation of PTF was performed on regions of the sky where PTF maintained a regular cadence for a prolonged duration.

Our study then used a sample of \lq normal\rq\ SNe Ia from PTF, with a definition following \citet{Betoule2014}, and contained 90 real SNe Ia: 84 spectroscopically confirmed, and 6 photometrically classified. This SN Ia sample was compared to the simulated objects (as a function of their SALT2 light-curve parameters and redshift), providing a weighting for each SN Ia event, and thus an overall rate. The statistical uncertainties were determined from our weighted sample. The systematics were calculated by performing several million realisations of the rate calculation, in each realisation drawing the SN Ia light-curve parameters from the covariance matrix of their SALT2 light-curve fit. The resulting 1-$\sigma$ confidence interval from this distribution of rate formed our systematic uncertainty.

We find the SN Ia rate in the local universe to be
\begin{equation*}
r_v=2.43\pm0.29\,\text{(stat)}_{-0.19}^{+0.33}\text{(sys)}\times10^{-5}\,\text{SNe\,yr}^{-1}\,\text{Mpc}^{-3}\, h_{70}^{3}.
\end{equation*}
This result is consistent with the expected rate evolution determined from other surveys.

Finally, we combined our result with a literature sample of SN Ia rates up to $z\sim2.5$. We used this cosmic rate evolution to constrain a simple delay-time model, $\propto\mathrm{t}^{-\beta}$. The delay-time distribution was convolved with three different star formation histories, and each fit to the rate evolution. We found that, regardless of our SFH assumption, $\beta{\sim}1$. This is consistent with a SN Ia progenitor channel dominated by the gravitational in-spiral of two white dwarfs. However, such a study of the delay-time distribution is now limited by uncertain high-redshift ($z>1$) SN Ia rates. 

Wide-fast-and-deep surveys, such as the Large Synoptic Survey Telescope \citep[LSST;][]{2009AAS...21346003I} are on the horizon. The latest survey predictions suggest that LSST could find $\approx50,000$ cosmically-useful SNe Ia (LSST Science Collaborations, in preparation)\footnote{The survey simulations are continually evolving, and the latest documentation can be found here: \url{https://github.com/LSSTScienceCollaborations/ObservingStrategy}} -- such a large sample size would significantly improve our constraints on the SN Ia rate evolution with redshift. Not only would this improve our understanding of the delay-time distribution and contributions from both SD and DD channels, but the methods presented in this paper are applicable for an LSST SNe Ia rate calculation. 

\section*{Acknowledgements}
We thank the anonymous referee for their useful comments.

We acknowledge support from EU/FP7 ERC grant number [615929]. We acknowledge the use of the IRIDIS High Performance Computing Facility, and associated support services, at the University of Southampton. This research used resources of the National Energy Research Scientific Computing Center, a DOE Office of Science User Facility supported by the Office of Science of the U.S. Department of Energy under Contract No. DE-AC02-05CH11231. Observations were obtained with the Samuel Oschin Telescope and the 60-inch Telescope at the Palomar Observatory as part of the Palomar Transient Factory project, a scientific collaboration between the California Institute of Technology, Columbia University, Las Cumbres Observatory, the Lawrence Berkeley National Laboratory, the National Energy Research Scientific Computing Center, the University of Oxford, and the Weizmann Institute of Science.









\appendix
\section{Spectroscopic Sample Objects}
\centering
\begin{table*}
\label{tab:spec_sample}
\caption{The spectroscopically confirmed sample of SNe Ia used to calculate the rate in Section~\ref{sec:sys_err}.}
\begin{tabular}{lrrrrrr}
\hline
PTF Name & R.A. (J2000) & Decl. (J2000) & z     & t$_0$ (MJD) & x$_1$ & M$_R$  \\\hline
PTF10cko    & 12:11:15.37  & +13:44:01.51  & 0.07  & 55266.93   & 0.61  & -18.83 \\
PTF10cwm    & 12:32:55.41  & +04:29:11.29  & 0.08  & 55271.97   & 0.61  & -18.63 \\
PTF10duy    & 13:57:11.07  & +40:09:47.89  & 0.079 & 55283.93   & 1.13  & -19.05 \\
PTF10duz    & 12:51:39.50  & +14:26:18.74  & 0.06  & 55285.78   & -0.26 & -18.88 \\
PTF10fkk    & 16:39:36.89  & +34:35:32.33  & 0.08  & 55308.21   & 1.22  & -19.07 \\
PTF10fps    & 13:29:25.06  & +11:47:46.50  & 0.022 & 55313.31   & -2.11 & -18.20 \\
PTF10fxl    & 16:52:47.55  & +51:03:44.87  & 0.03  & 55318.44   & -0.12 & -18.63 \\
PTF10gjx    & 12:24:54.17  & +20:05:04.24  & 0.076 & 55326.49   & 0.51  & -18.89 \\
PTF10glo    & 12:33:05.76  & +06:33:22.66  & 0.075 & 55322.89   & 1.14  & -18.96 \\
PTF10gmg    & 16:24:58.50  & +51:02:21.53  & 0.063 & 55324.98   & 0.95  & -19.05 \\
PTF10gmj    & 16:43:09.09  & +66:20:51.46  & 0.082 & 55315.58   & -0.29 & -18.74 \\
PTF10gnj    & 12:14:41.78  & +10:57:33.15  & 0.08  & 55318.24   & 1.26  & -18.05 \\
PTF10hdn    & 14:52:24.64  & +47:28:35.23  & 0.07  & 55343.39   & 0.45  & -18.86 \\
PTF10hmc    & 16:32:45.81  & +35:04:07.62  & 0.073 & 55342.35   & 0.14  & -19.28 \\
PTF10hml    & 13:19:49.69  & +41:59:01.60  & 0.054 & 55353.0    & 0.54  & -18.99 \\
PTF10hmv    & 12:11:32.99  & +47:16:29.79  & 0.032 & 55352.09   & 1.17  & -18.41 \\
PTF10hoz    & 16:44:52.69  & +52:37:01.78  & 0.06  & 55348.3    & 0.18  & -18.69 \\
PTF10hrw    & 17:44:15.47  & +52:08:57.88  & 0.049 & 55349.75   & -0.42 & -18.79 \\
PTF10ifj    & 14:19:21.93  & +34:21:14.30  & 0.076 & 55354.66   & 1.21  & -19.03 \\
PTF10inf    & 16:43:15.06  & +32:40:27.77  & 0.05  & 55361.09   & 0.13  & -18.78 \\
PTF10iyc    & 17:09:21.83  & +44:23:35.90  & 0.055 & 55362.09   & 0.66  & -18.88 \\
PTF10jdw    & 15:42:00.31  & +47:35:37.96  & 0.077 & 55366.91   & -0.56 & -18.84 \\
PTF10jtp    & 17:10:58.47  & +39:28:28.28  & 0.067 & 55365.03   & -0.5  & -18.58 \\
PTF10jwx    & 13:42:39.30  & +56:27:44.36  & 0.068 & 55368.62   & -0.85 & -18.97 \\
PTF10kdg    & 13:17:13.70  & +44:08:39.45  & 0.062 & 55370.68   & -0.45 & -18.42 \\
PTF10kiw    & 14:46:18.08  & +47:12:26.22  & 0.069 & 55369.44   & 0.69  & -18.17 \\
PTF10lot    & 15:38:39.73  & +41:00:19.88  & 0.025 & 55381.93   & 0.44  & -18.97 \\
PTF10lxp    & 14:23:56.80  & +55:43:44.73  & 0.088 & 55381.96   & 0.2   & -19.24 \\
PTF10lya    & 23:41:31.67  & +14:07:25.14  & 0.064 & 55380.76   & 0.09  & -18.82 \\
PTF10mbk    & 14:17:04.79  & +71:47:23.54  & 0.065 & 55381.7    & 0.59  & -19.09 \\
PTF10mwb    & 17:17:49.97  & +40:52:52.06  & 0.03  & 55391.34   & -0.58 & -18.61 \\
PTF10ncu    & 17:49:01.12  & +68:06:18.51  & 0.07  & 55381.0    & -2.31 & -18.14 \\
PTF10ndc    & 17:19:50.18  & +28:41:57.46  & 0.082 & 55391.0    & 0.78  & -19.08 \\
PTF10nlg    & 16:50:34.48  & +60:16:34.95  & 0.05  & 55392.61   & -0.38 & -18.15 \\
PTF10nvh    & 21:32:02.35  & +08:59:35.71  & 0.068 & 55394.24   & 0.39  & -18.82 \\
PTF10otc    & 16:59:16.38  & +68:10:55.38  & 0.054 & 55390.37   & -2.95 & -18.51 \\
PTF10pvi    & 22:02:02.32  & +14:32:10.39  & 0.08  & 55410.83   & 0.23  & -18.94 \\
PTF10qsc    & 21:34:21.21  & -05:03:44.39  & 0.088 & 55422.87   & 1.01  & -19.03 \\
PTF10qwg    & 02:42:09.88  & +02:26:52.48  & 0.071 & 55425.38   & -0.41 & -18.88 \\
PTF10qyx    & 02:27:12.06  & -04:32:04.79  & 0.063 & 55426.98   & -1.39 & -18.69 \\
PTF10rbp    & 01:16:38.06  & -01:49:23.60  & 0.079 & 55431.38   & 0.64  & -18.85 \\
PTF10rhi    & 23:49:43.77  & +13:02:33.27  & 0.085 & 55425.57   & 0.03  & -18.93 \\
PTF10tce    & 23:19:10.36  & +09:11:54.23  & 0.041 & 55442.89   & 0.83  & -18.85 \\
PTF10tqy    & 00:42:41.52  & +24:45:14.42  & 0.045 & 55444.61   & -1.76 & -18.74 \\
PTF10trp    & 21:28:07.98  & +09:51:13.18  & 0.049 & 55450.76   & 1.55  & -18.16 \\
PTF10trs    & 00:15:20.64  & +17:31:59.38  & 0.073 & 55442.31   & -1.59 & -18.87 \\
PTF10ubm    & 00:01:59.28  & +21:49:29.83  & 0.07  & 55455.7    & 0.77  & -19.23 \\
PTF10ucl    & 22:06:04.85  & +15:30:34.47  & 0.08  & 55444.95   & -0.72 & -19.00 \\
PTF10ufj    & 02:25:39.13  & +24:45:53.16  & 0.073 & 55457.36   & -0.04 & -18.65 \\
PTF10vfo    & 01:15:06.67  & +24:36:17.77  & 0.088 & 55459.24   & -0.32 & -18.86 \\
PTF10viq    & 22:20:19.97  & +17:03:22.23  & 0.031 & 55452.94   & 0.69  & -18.81 \\
PTF10wnm    & 00:22:03.61  & +27:02:26.25  & 0.066 & 55477.67   & 0.43  & -18.94 \\
PTF10wnq    & 00:49:09.82  & +32:08:19.46  & 0.07  & 55473.45   & -0.46 & -18.97 \\
PTF10wof    & 23:32:41.84  & +15:21:31.71  & 0.053 & 55475.08   & 0.05  & -18.68 \\
PTF10wyq    & 01:18:24.89  & +19:26:46.20  & 0.08  & 55480.18   & 0.47  & -18.66 \\
PTF10xyl    & 00:20:24.86  & +05:42:05.75  & 0.056 & 55477.95   & -2.66 & -18.21 \\
PTF11bas    & 13:16:47.94  & +43:31:13.27  & 0.085 & 55641.06   & 0.24  & -18.75 \\
PTF11biv    & 17:08:33.90  & +22:04:15.95  & 0.048 & 55639.64   & 0.88  & -18.07 \\
PTF11bnx    & 16:30:19.32  & +21:05:07.27  & 0.06  & 55653.87   & -0.07 & -18.65 \\
PTF11bui    & 13:12:56.40  & +47:27:12.46  & 0.028 & 55675.79   & 0.81  & -18.83 \\
PTF11cao    & 16:18:47.96  & +25:11:16.47  & 0.04  & 55671.11   & 0.28  & -18.60 \\
PTF11dif    & 14:15:52.07  & +35:44:23.95  & 0.059 & 55703.67   & 0.62  & -18.93 \\
PTF12bok    & 12:13:37.15  & +46:29:01.14  & 0.025 & 56015.63   & -0.48 & -18.81 \\
\hline\end{tabular}
\end{table*}

\centering
\begin{table*}
\contcaption{}
\begin{tabular}{lrrrrrr}
\hline
PTF Name & R.A. (J2000) & Decl. (J2000) & z     & t$_0$ (JD) & x$_1$ & M$_R$  \\\hline
PTF12ccz    & 16:03:42.04  & +36:59:44.22  & 0.041 & 56018.76   & -2.94 & -18.14 \\
PTF12cjg    & 13:41:18.26  & +55:27:07.85  & 0.067 & 56025.53   & -0.39 & -18.80 \\
PTF12cks    & 14:22:44.18  & +34:15:16.25  & 0.063 & 56028.94   & 0.99  & -19.01 \\
PTF12cnl    & 13:11:07.30  & +39:04:55.62  & 0.047 & 56034.64   & 1.25  & -18.81 \\
PTF12csi    & 16:47:13.82  & +33:18:20.53  & 0.053 & 56031.48   & 0.76  & -18.03 \\
PTF12dco    & 15:20:31.73  & +59:11:44.64  & 0.075 & 56040.5    & 0.42  & -18.84 \\
PTF12dhb    & 16:16:57.43  & +49:41:50.49  & 0.056 & 56042.06   & 0.36  & -18.88 \\
PTF12dhl    & 13:22:55.63  & +52:14:00.48  & 0.057 & 56041.42   & -2.79 & -18.78 \\
PTF12dxm    & 13:53:26.48  & +43:54:48.61  & 0.063 & 56054.22   & -1.67 & -18.53 \\
PTF12eac    & 16:53:22.58  & +36:16:23.29  & 0.088 & 56056.25   & 1.85  & -18.39 \\
PTF12ecm    & 15:56:21.31  & +36:32:13.78  & 0.066 & 56068.35   & 0.53  & -18.83 \\
PTF12ecr    & 14:35:46.52  & +45:11:19.67  & 0.069 & 56067.84   & 0.64  & -18.72 \\
PTF12fuu    & 15:04:40.39  & +06:04:21.01  & 0.035 & 56112.27   & -0.01 & -18.18 \\
PTF12gaz    & 15:37:39.90  & +06:36:58.02  & 0.071 & 56113.69   & -0.45 & -18.84 \\
PTF12gdq    & 15:11:35.31  & +09:42:34.04  & 0.033 & 56116.99   & -0.71 & -18.65 \\
PTF12ggb    & 15:38:25.64  & +31:32:08.83  & 0.06  & 56117.02   & 0.58  & -18.86 \\
PTF12gkn    & 15:25:39.83  & +09:39:05.66  & 0.077 & 56120.42   & 1.03  & -18.82 \\
PTF12gmv    & 02:45:01.44  & -00:43:48.46  & 0.054 & 56115.02   & -0.04 & -18.65 \\
PTF12hmx    & 22:47:09.01  & +00:10:27.97  & 0.085 & 56150.8    & -0.76 & -18.53 \\
PTF12iiq    & 02:50:07.76  & -00:15:54.45  & 0.029 & 56182.74   & -1.29 & -18.63 \\
PTF12ikt    & 01:14:43.13  & +00:17:07.11  & 0.044 & 56187.79   & -0.22 & -18.85 \\
\hline
\end{tabular}
\end{table*}

\bsp	
\label{lastpage}
\end{document}